\documentclass{nature}
\usepackage{graphicx}
\usepackage{color}
\usepackage{epsfig}
\title{Cooperative concurrence of 4$f$ and 3$d$ flat bands in kagome heavy-fermion metal YbCr$_{6}$Ge$_{6}$}
\author{Wenxin Lv$^{1,2,\dag}$, Pengcheng Ma$^{3,\dag}$, Tianqi Wang$^{3}$, Shangjie Tian$^{4,1,2}$, Ying Ma$^{5}$, Shouguo Wang$^{4}$, Xiao Zhang$^{5,*}$, Zhonghao Liu$^{3,*}$, and Hechang Lei$^{1,2,*}$}

\begin{document}
\maketitle

\begin{affiliations}
 \item School of Physics and Beijing Key Laboratory of Optoelectronic Functional Materials $\&$ MicroNano Devices, Renmin University of China, Beijing 100872, China
 \item Key Laboratory of Quantum State Construction and Manipulation (Ministry of Education), Renmin University of China, Beijing 100872, China
 \item Institute of High-Pressure Physics and School of Physical Science and Technology, Ningbo University, Ningbo 315211, China
 \item School of Materials Science and Engineering, Anhui University, Hefei 230601, China
 \item State Key Laboratory of Information Photonics and Optical Communications \& School of Physical Science and Technology, Beijing University of Posts and Telecommunications, Beijing 100876, China
 \end{affiliations}

\vspace*{3cm}
\leftline{$^\dag$These authors contributed equally to this work.}
\leftline{$^*$Corresponding authors: Z.X. (zhangxiaobupt@bupt.edu.cn); Z.H.L. (liuzhonghao@nbu.edu.cn); }
\leftline{H.C.L. (hlei@ruc.edu.cn)}

\newpage

\begin{abstract}
Flat-band (FB) systems originating from special lattice geometry like in kagome metals as well as localized orbitals in the materials such as heavy-fermion (HF) compounds have induced intensive interest due to their band topology and strong electron correlation effects, leading to emergent quantum states of matter.
However, the question of how these two distinct FBs coexist and interact remains unsettled.
Here, we report that YbCr$_{6}$Ge$_{6}$ hosting both Cr-kagome lattice and Yb-$4f$ electrons exhibits HF behaviors and a robust antiferromagnetic ground state with transition temperature $T_{N}$ = 3 K, significantly higher than other similar kagome metals with Yb ions.
Angle-resolved photoemission spectroscopy measurements reveal the coexistence of FBs originating from both Cr-kagome lattice and localized Yb-4$f$ electrons near Fermi energy level $E_{\rm F}$. 
More importantly, the clear spectroscopic signatures of a hybridization of Yb-4$f$ FB with kagome-lattice-derived conduction bands and the high density of states of Cr-kagome FB  near $E_{\rm F}$ provide the underlying microscopic mechanisms of HF behaviors and enhanced antiferromagnetism in YbCr$_{6}$Ge$_{6}$. 
Our findings demonstrate that the novel kagome HF metals can not only host the cooperative coexistence of two different types of FBs, but also provide a paradigm material platform to explore the exotic correlated topological quantum phenomena.
\end{abstract}

The exploration of novel quantum states of matter is one of central topics in condensed matter physics. The systems with flat bands (FBs) emerge as a particularly fertile ground because the FBs not only suppress the electronic kinetic energy dramatically and enhance electron correlations significantly, but also process the nontrivial band topology \cite{Checkelsky}. 
The FBs can drive electronic systems towards a variety of strongly correlated states, including ferromagnetism, unconventional superconductivity and fractional Chern insulators \cite{Mielke,Ko,Tang}.
To date, several conceptually distinct strategies have been employed to engineer such FBs.

The first paradigm is lattice-geometry-driven FBs, epitomized by the two-dimensional (2D) kagome lattice. Such kind of lattice composed of corner-sharing triangles inherently hosts a destructive interference of electron wavefunctions that leads to a FB in its single-particle electronic structure, accompanied by Dirac cones and van Hove singularities \cite{YeL,YinJX1}. 
This unique electronic structure has rendered kagome metals like the T$_{m}$Sn$_{n}$ (T = Mn, Fe, Co), RET$_{6}$X$_{6}$ (RE = rare-earth elements, X = Ge, Sn), and AV$_{3}$Sb$_{5}$ family (A = K, Rb, Cs)  as celebrated systems for investigating the intricate interplay between magnetism, charge order, superconductivity and band topology \cite{YeL,LiuZ,Ortiz1,Ortiz2,YinQW,YinJX2,WangQ,LiM,Arachchige}. 
In these materials, the low-energy physics is primarily dictated by the quantum geometry of the lattice and the correlations among itinerant $d$-electrons. 
A second fundamentally different route to construct FBs is utilizing the many-body effects, which has been realized in heavy-fermion (HF) systems \cite{Wirth}.
In these materials, typically containing lanthanide or actinide elements, the hybridization between localized $f$-electrons and dispersive itinerant electrons--the Kondo effect--gives rise to narrow, correlation-induced HF FBs with resonance peak of density of states (DOS) at the Fermi energy level ($E_{\rm F}$). 
These emergent HF FBs, directly visualized by angle-resolved photoemission spectroscopy (ARPES) \cite{Shim,ChenQY}, don't originate from the lattice geometry but instead are pure manifestation of strong electron correlations, hosting heavy quasiparticles with effective masses up to a thousand times that of a free electron.

Despite the remarkable progress have made in these two parallel fields, the physics of lattice-geometry-driven kagome FBs and electron-correlation-driven $f$-electron FBs have remained largely disconnected. 
This raises an unexplored question: what novel quantum phenomena would emerge when these two distinct types of FBs coexist and interact with each other in a single material? 
%Can the geometric frustration of a kagome lattice modulate the formation of the Kondo resonance? Conversely, how does the strong correlation from f-electrons reconstruct the pristine electronic structure of the Kagome lattice? 
%Answering these questions requires the discovery of a new material that synergistically integrates both key ingredients.
In this work, we carried out a detailed study on YbCr$_{6}$Ge$_{6}$ hosting both kagome lattice and Yb-$4f$ electrons, which is a paradigm system to answer the above question. 
YbCr$_{6}$Ge$_{6}$ exhibits an antiferromagnetic (AFM) ground state of Yb ions with transition temperature $T_{N}\sim $ 3 K, which is much larger than that in YbV$_{6}$Sn$_{6}$ \cite{GuoKZ}. In addition, it exhibits the HF features above $T_{N}$.
ARPES results further confirm that the Yb-4$f$ FBs coexist with the Cr-3$d$ FB originating from the kagome lattice near the $E_{\rm F}$ in YbCr$_{6}$Ge$_{6}$.
The Kondo hybridization of Yb-4$f$ FBs with itinerant electrons leads to the HF behaviors while the high DOS of Cr-3$d$ FB enhances the $T_{N}$ of AFM ground state.

\section*{Results}

\textbf{Structure and transport behaviors in YbCr$_{6}$Ge$_{6}$.} 
LnCr$_{6}$Ge$_{6}$  (Ln = Yb and Lu) single crystals were grown by using Sn flux. The detailed structural and composition characterizations are shown in Supplementary Figure 1. 
LnCr$_{6}$Ge$_{6}$ has a hexagonal crystal structure (space group $P6/mmm$, No. 191) \cite{Romaka}.
It consists of Ln-Ge layer with Ln triangle lattice and Cr-Ge-Ge-Ge-Cr slab with double Cr kagome lattice stacking along $c$ axis alternatively, as shown in Figs. 1a and 1b.
Figures 1c and 1d depict the temperature dependence of magnetic susceptibility $\chi(T)$ under the field of 1 T for $H\Vert c$ and $H\Vert ab$, respectively. 
The $\chi(T)$ curve shows obvious easy-axis anisotropy in the temperature range from 2 to 300 K.
The $\chi_{ab}(T)$ curve exhibits a wide hump near 50 K before the quick increase at lower temperature. In contrast, the $\chi_{c}(T)$ curve doesn't show such anomaly. 
%This anisotropic behavior of $\chi(T)$ curves of YbCr$_{6}$Ge$_{6}$ is different from that in YbV$_{6}$Sn$_{6}$ \cite{GuoKZ}, where the anomaly appears in $\chi_{c}(T)$ curve.
For both field directions, the high-temperature $\chi(T)$ curves can be fitted well using the Curie-Weiss (CW) law, $\chi (T)=\chi_{0}+C/(T - \theta)$, where $C$ is Curie constant and $\theta$ is Weiss temperature. The linear behaviors of $(\chi (T)-\chi_{0})^{-1}$ curves at high temperature (blue squares in Figs. 1c and 1d) also indicate the validation of CW law.
From the fitted $C$, the calculated effective moment $\mu_{{\rm eff},c}$ is 4.81(2) $\mu_{\rm B}$/f.u. and $\theta_{c}$ = -40.0(4) K for $H\Vert c$ and  $\mu_{{\rm eff},ab}$ = 5.18(6) $\mu_{\rm B}$/f.u. and $\theta_{ab}$ = -69.5(4) K for $H\Vert ab$.
The fitted $\theta$ values for both field directions are negative, implying the AFM interactions in this material. 
%The larger $\theta_{ab}$ than $\theta_{c}$ reflects large anisotropy of interaction and the possible dominant interaction along the $ab$ plane.
Both values of $\mu_{{\rm eff}}$ are larger than the theoretical value of Yb$^{3+}$ ($\mu_{\rm eff}$ = 4.54 $\mu_{\rm B}$/f.u.), suggesting the valent state of Yb at high temperature is mainly +3 and Cr ions may have local moments. In order to confirm this conjecture, we measured the $\chi (T)$ curves of LuCr$_{6}$Ge$_{6}$ (Supplementary Fig. 2). 
The fitted $\mu_{{\rm eff},c}$ and $\mu_{{\rm eff},ab}$ is 0.88(6) and 0.85(8) $\mu_{\rm B}$/f.u., respectively, when the $\theta_{c}$  and $\theta_{ab}$ are -40.4(8) and -57.8(9) K. It further indicates Cr ions should have local moments and AFM interactions.
As shown in the insets of Figs. 1c and 1d,  the $\chi_{ab}(T)$ curve shows a drop below 3 K at 0.01 T while the $\chi_{c}(T)$ curve only exhibits a kink at same field. It indicates that there is an AFM transition with spin direction along the $ab$ plane in YbCr$_{6}$Ge$_{6}$.
With the increase of magnetic field, the drop/kink on $\chi(T)$ curve are suppressed gradually to lower temperature, confirming the nature of transition is AFM further.

As shown in Fig. 1e, the field-dependent magnetization $M(\mu_{0}H)$ for $H\Vert c$ is much larger than that for $H\Vert ab$ at 2 K, which could be due to the spin canting along the $c$ axis. Moreover, for $H\Vert c$, the $M(\mu_{0}H)$ curve shows a linear increasing trend in the range from 0 to about 0.9 T, then gradually saturates and reaches 1.40 $\mu_{\rm B}$/f.u. at 7 T, which is much lower than the saturation magnetic moment of Yb$^{3+}$ of 4.0 $\mu_{\rm B}$/f.u.. It implies that there may be a magnetization plateau phase in YbCr$_{6}$Ge$_{6}$ with Yb triangle lattice. 
%When the temperature is 10 K, the inflection point in the $M(H)$ curve in both directions disappears.
When compared to LuCr$_{6}$Ge$_{6}$, the $ab$-plane resistivity $\rho_{ab}(T)$ curve of YbCr$_{6}$Ge$_{6}$ shows a hump near 90 K and a fast decrease at around 3 K (Fig. 1f). 
The latter should be related to the suppression of spin disorder scattering below the AFM transition. 
The inset of Fig. 1f shows the difference in resistivity $\Delta\rho_{ab}(T) = \rho_{ab, \rm {Yb}}(T) - \rho_{ab, \rm {Lu}}(T)$ in order to estimate the contribution of $4f$ electrons.
The $\Delta\rho_{ab}(T)$ displays a maximum at $\sim$ 90 K while the low-temperature $\Delta\rho_{ab}(T)$ decreases remarkably in a magnetic field, suggesting that the resistivity hump of YbCr$_{6}$Ge$_{6}$ at high temperature could be due to the appearance of Kondo coherence. 

For the specific heat results of YbCr$_{6}$Ge$_{6}$ and LuCr$_{6}$Ge$_{6}$ at zero field (Fig. 2a), both of them at high temperature are close to the Dulong-Petit limit ($3NR$, blue line), where $N$ (= 13) is the atomic number and $R$ (= 8.314 J mol$^{-1}$ K$^{-1}$) is the ideal gas constant. 
For LuCr$_{6}$Ge$_{6}$, the low-temperature specific heat can be fitted using the relation $C_{p}/T = \gamma + \beta T^{2}$, where $\gamma$ is electronic specific heat coefficient and $\beta$ is lattice specific heat coefficient. The fitted $\gamma$ is 73.4(4) mJ mol$^{-1}$ K$^{-2}$ when the values of $\beta$ are 0.24(1) mJ mol$^{-1}$ K$^{-4}$. Correspondingly, the calculated Debye temperature $\Theta_{D}$ is 475(6) K using the relationship $\Theta_{\rm D} = (\frac{12 \pi^{4} NR}{5 \beta})^{1/3}$. 
The $\gamma$ value of LuCr$_{6}$Ge$_{6}$ is much larger than that of LuV$_{6}$Sn$_{6}$ (17.4 mJ mol$^{-1}$ K$^{-2}$) \cite{GuoKZ}, which is likely to reflect that the electrons in Cr-based kagome lattice have stronger correlation effects than that in V-based kagome lattice. 
%Furthermore, the larger $\gamma$ value of LuCr$_{6}$Ge$_{6}$ also indicates the strong-correlation feature of Cr-Kagome lattice.
Figure 2b displays the specific heat data of YbCr$_{6}$Ge$_{6}$ in different magnetic fields at the low-temperature region. 
The zero-field $C_{p}/T$ curve shows an upward trend when $T<$ 8 K and a $\lambda$-type peak appears at 3 K, indicating a long-range magnetic ordering state emerges. The ordering temperature is consistent with that determined from the $\chi (T)$ curves. 
For zero-field data, we fit the low-temperature specific heat from 9 K to 14 K. The obtained $\gamma$ value is 203(8) mJ mol$^{-1}$ K$^{-2}$, significantly larger than that of LuCr$_{6}$Ge$_{6}$. This large $\gamma$ value suggests the HF feature of YbCr$_{6}$Ge$_{6}$ when $T > T_{N}$.
When the magnetic field ($\leq$ 1 T) is applied along the $c$-axis, the $\lambda$-type peak moves to lower temperatures continuously, reflecting the AFM feature of long-range ordering in YbCr$_{6}$Ge$_{6}$. At field higher than 1 T, the $\lambda$-type peak turns into a broad peak, which moves to higher temperatures with increasing the field. 
Such behavior can be observed more clearly in the magnetic specific heat $C_{m}(T)$ data (Fig. 2c), which are obtained by subtracting the specific heat of non-magnetic analogue LuCr$_{6}$Ge$_{6}$.
The magnetic entropy $S_{m}(T)$ is calculated by integrating $C_{m}(T)$ (Fig. 2d). It is noteworthy that due to the experimental limitation where measurements only extend down to 2 K, the calculated $S_{m}(T)$ may be subject to an underestimation. 
It is found that the $S_{m}(T)$ value is about 25 $\%$ of $R$ln(2) at $T_{\rm N}$. Then, it increases gradually and reaches a value of $R$ln(2) and $R$ln(4) at $\sim$ 35 K and $\sim$ 80 K, respectively. At $T>$ 100 K, the $S_{m}(T)$ saturates to about 14 J mol$^{-1}$ K$^{-1}$, which is close to the value of $R$ln(6) ($\sim$ 14.9 J mol$^{-1}$ K$^{-1}$).
This result suggests that the Yb ions in YbCr$_{6}$Ge$_{6}$ possess a Kramers doublet as their ground state at low temperature.

\textbf{Experimental electronic structure observed by ARPES measurements.}
To uncover the microscopic origin of the enhanced magnetic and electronic correlations in YbCr$_6$Ge$_6$ and to directly probe the interplay between the Cr-derived kagome bands and the Yb-$4f$ states, we performed systematic ARPES measurements on both surface terminations of YbCr$_6$Ge$_6$ and compared them with isostructural LuCr$_6$Ge$_6$. 
Figures 3a and 3b display the Fermi surfaces (FSs) and core-level spectra obtained from the Cr- and YbGe-terminated surfaces. Although the two terminations exhibit similar FS topology, their spectral intensities differ substantially due to the much larger DOS contributed by Yb-4$f$ electrons near $E_{\rm F}$ on the YbGe termination. The two terminations can be unambiguously identified from their core-level spectra. On the Cr-terminated surface, the peaks derived from Yb-$4f$ states are relatively simple and weak. In contrast, on the YbGe-terminated surface, multiple well-separated $4f$ peaks originating from different Yb layers are observed around -12 to -5 eV, as well as from -2 eV up to $E_{\rm F}$, consistent with previous reports \cite{Kummer,Lou}.

Over a wide binding-energy range shown in Fig. 3c, the canonical kagome features (Fig. 3d) previously reported in YCr$_6$Ge$_6$ \cite{YangTY} are visible on the Cr termination, together with FBs near $E_{\rm F}$ and at -1.31 eV, the latter corresponding to the Yb 4$f_{7/2}$ state. In contrast, on the YbGe termination, four nearly equally energy-spaced ($\sim 0.66$ eV) $4f$-derived FBs are clearly resolved in both the intensity maps and the integrated energy distribution curves (EDCs), and these FBs intersect the kagome-derived conduction bands. When resonant photoemission measurements are performed using 186 eV photons corresponding to the Yb $4d$--$4f$ resonance, the four main FBs (the red solid lines) are significantly enhanced, and an additional set of three weak, nearly equally spaced sub-FBs (the blue dashed lines) emerges at energies midway between the main FBs. These weaker features likely originate from Yb atoms occupying different crystallographic environments in the sample \cite{Weiland}. Figure 3e compares the terminated-dependent band dispersions along the $\bar{\Gamma}$--$\bar{K}$ direction of YbCr$_6$Ge$_6$ with those of LuCr$_6$Ge$_6$. The Cr-terminated band structure of YbCr$_6$Ge$_6$ is nearly identical to that of LuCr$_6$Ge$_6$, except for the presence of an additional FB around -1.31 eV derived from the Yb 4$f_{7/2}$ state. In contrast, the YbGe-terminated surface exhibits clear crossings between the $4f$-derived FBs and the kagome conduction band, giving rise to pronounced $c$--$f$ hybridized states, as schematically illustrated in Fig. 3f, which constitute a hallmark of Kondo hybridization.

To directly resolve the $c$--$f$ hybridization and examine its temperature evolution, we performed temperature-dependent ARPES measurements along the $\bar{M}$--$\bar{\Gamma}$--$\bar{M}$ direction on the $k_z \sim$ 0 plane of the YbGe-terminated surface, where the band-crossing features are more clearly resolved than along the $\bar{K}$--$\bar{\Gamma}$--$\bar{K}$ direction. Figure 4a presents the spectra measured at 150 K and 20 K. At low temperature, all four $4f$-derived FBs sharpen markedly, and hybridization gaps open at their crossings with the conduction band near -0.3 eV and -0.6 eV below $E_{\rm F}$, as highlighted by the red rectangles, which is clearly distinct from that at 150~K, as well as from that of LuCr$_6$Ge$_6$ shown in Fig. 4b. A zoomed-in view of the near-$E_{\rm F}$ region at 150, 115, 85, and 20 K is presented in Fig. 4c, where the hybridization feature near $E_{\rm F}$ and -0.3 eV becomes prominent at low temperature, as further corroborated by the enlarged images and EDCs in Fig. 4d.

To quantitatively track the temperature dependence of these features, we extracted EDCs at three representative momentum positions (blue square, triangle, and circle in Fig. 4c) and fitted them using a polynomial background combined with Lorentzian peak functions. At the square position, the peak 1 near $E_{\rm F}$ remains essentially temperature independent, whereas the peak 2 near -0.3 eV shifts upward toward $E_{\rm F}$ with increasing temperature, demonstrating that the hybridization gap closes at approximately 120 K. This characteristic temperature is higher than the Kondo coherence temperature inferred from transport measurements (Fig. 1f), a behavior commonly observed in Kondo lattice systems. 
At the triangle position, the EDC peak 3 near $E_{\rm F}$ exhibit negligible temperature dependence, consistent with the kagome-derived conduction band being largely insensitive to temperature. 
At the circle position, the EDC peak 4 gradually move toward $E_{\rm F}$ upon warming, following the same trend as the -0.3 eV peak 2 at the square position and confirming their origin in the $4f$-derived states. At high temperatures, the $4f$-derived hybridized bands near $E_{\rm F}$ lose coherence and eventually vanish, leaving behind a faint shadow of the kagome FB similar to that observed in YCr$_6$Ge$_6$ \cite{YangTY}.
Because the fitting procedure explicitly accounts for thermal broadening, the observed peak shifts reflect genuine changes in the electronic structure rather than trivial thermal effects. In addition, the temperature evolution of peak 2 and peak 4 exhibits a noticeable kink around 70 -- 85 K, consistent with the Kondo coherence temperature inferred from transport measurements (Fig. 1f). 
%These ARPES results demonstrate that YbCr$_6$Ge$_6$ hosts two distinct yet cooperative flat-band components: a kagome-geometry-imposed FB present in both LuCr$_6$Ge$_6$ and YbCr$_6$Ge$_6$, and a correlation-driven $4f$ FB unique to YbCr$_6$Ge$_6$ that emerges from Kondo hybridization at low temperature. 

\section*{Discussion}
When compared to YbV$_{6}$Sn$_{6}$ ($T_{N} \sim$ 0.4 K) \cite{GuoKZ}, the $T_{N}$ of YbCr$_{6}$Ge$_{6}$ is significantly enhanced ($\sim$ 3 K). 
It implies that the existence of local moments of Cr ions in kagome plane compared to the non-magnetic V ions may have some influence on long-range ordering of Yb-4$f$ electrons. 
On the other hand, the existence of Cr-kagome FB dramatically increases the DOS near $E_{\rm F}$, $N(E_{\rm F})$. According to the Ruderman-Kittel-Kasuya-Yosida (RKKY) theory, the large $N(E_{\rm F})$ serves as a highly efficient medium for the exchange interaction, naturally leading to a much stronger magnetic coupling and a significant enhancement of $T_{N}$.
%At a higher energy scale, the hybridization gap observed in ARPES begins to close around 120 K, a temperature slightly higher than the Kondo coherence temperature reflected in susceptibility [Fig. 2(b)] and resistivity [Fig. 2(d)], which is also consistent with previous reports[xxx]. 
The coexistence and interplay of these two types of FBs--one lattice-geometry-driven and one electron-correlation-driven--provide a natural microscopic explanation for the synergistic enhancement of magnetic and electronic correlations revealed by the transport measurements. 

In summary, we demonstrate that YbCr$_{6}$Ge$_{6}$ with Cr kagome lattice exhibit HF features of Yb-4$f$ electrons with the ground state of AFM order below $T_{N}$ = 3 K.
Further experimental results reveal that YbCr$_{6}$Ge$_{6}$ hosts two distinct yet cooperative FB components: a kagome-geometry-imposed one arising from destructive interference in the Cr kagome lattice, and an electron-correlation-driven Yb-4$f$ FB associated with Kondo hybridization at low temperature.
Present study establishes a new class of quantum states of matter where topological band structure and strong $f$-electron correlations can be studied in synergy. It opens a new frontier for exploring correlated topological phenomena and quantum critical behaviors.

%Note added.-After we have submitted our work, we notice that there is a preprints \cite{LiuE} reporting the study on the electronic structure and AHE of Co$_{3}$Sn$_{2}$S$_{2}$ single crystal, which shares consistent conclusions to parts of ours.

\begin{methods}

\noindent\textbf{Single crystal growth.} The single crystals of YbCr$_{6}$Ge$_{6}$ and LuCr$_{6}$Ge$_{6}$ were grown by using Sn-flux methods. The starting materials Yb (stick, 99.99\%), Lu (stick, 99.99 \%), Cr (pieces, 99.9999 \%), Ge (pieces, 99.999 \%) and Sn (grains, 99.999 \%) with the molar ratios Yb : Cr : Ge : Sn = 1.5 : 1.167 : 6 : 20 and Lu : Cr : Ge : Sn = 1 : 3 : 6 : 20 were mixed and placed in a corundum Canfield crucible set, which is effective at preventing samples from contacting the silica wool, and then sealed in a quartz tube with backfilled Ar gas ($\sim$ 150 Torr). The sealed quartz tube was put into a box furnace and heated to 1323 K in 24 h. After keeping at 1323 K for 10 h, the quartz tube was rapidly cooled at a rate of 5 K/h to 1223 K in order to prevent phase change of quartz tube. Then, it was cooled down to 973 K at 2.5 K/h and the Sn flux is removed by centrifugation. The shiny crystals with typical size about 1 $\times$ 1 $\times$ 0.1 mm$^{3}$ can be obtained. The samples were soaked in hydrochloric acid for a short time to remove residual Sn flux on the surface.

\noindent\textbf{Structural and composition characterizations.} X-ray diffraction (XRD) patterns of YbCr$_{6}$Ge$_{6}$ and LuCr$_{6}$Ge$_{6}$ were measured using a Bruker D8 X-ray diffractometer with Cu $\kappa_{\alpha}$ ($\lambda$ = 1.5418 \AA) radiation at room temperature. The elemental analysis was performed using an energy-dispersive X-ray spectroscopy (EDX) analysis in a FEI Nano 450 scanning electron microscope.

\noindent\textbf{Physical properties measurements.} Magnetization measurements were performed using Quantum Design magnetic property measurement system (MPMS3). Electrical transport and heat capacity measurements were performed using physical property measurement system (PPMS-14T). Electrical transport measurements were conducted using a standard four-probe configuration with the current parallel to the $ab$ plane and magnetic field along the $c$ axis.

\noindent\textbf{Angle-resolved photoemission spectroscopy (ARPES) measurements.} The ARPES measurements were conducted over a wide range of photon energies at the 03U and 09U endstations of the Shanghai Synchrotron Radiation Facility (SSRF). The $\mu$-ARPES setup featured a beam size smaller than 20 $\times$ 20 $\mu$m$^{2}$. The energy and momentum resolutions were maintained at better than approximately 20 meV and 0.02 \AA$^{-1}$, respectively. Samples smaller than 1 $\times$ 1 mm$^{2}$ were cleaved in situ to produce flat, mirror-like (00$l$) surfaces. During temperature-dependent measurements, the pressure was maintained at a level below 2 $\times$ 10$^{-10}$ Torr.

\end{methods}

\section*{References}

%\bibliography{YbCr6Ge6}

\begin{addendum}

\item This work is supported by the National Key R\&D Program of China (Grant Nos. 2022YFA1403800 and 2023YFA1406500), the National Natural Science Foundation of China (Grant Nos. 12274459 and 12222413), the Natural Science Foundation of Shanghai (Grants No. 23ZR1482200), and the Natural Science Foundation of Ningbo (Grants No. 2024J019), the Innovation Yongjiang 2035 Key R\&D Programme-International Sci-tech Cooperation Projects, the funding of Ningbo Yongjiang Talent Program and the Mechanics Interdisciplinary Fund for Outstanding Young Scholars of Ningbo University (Grants No. LJ2024003). We thank the SSRF of BL03U (31124.02.SSRF.BL03U) for the assistance with ARPES measurements.

\item[Author contributions] H.C.L., provided strategy and advice for the research; 
W.X.L., S.J. T. and Y. M. performed the crystal growth, structural and physical properties characterizations and fundamental data analysis with the assistance of S.G.W., X.Z., and H.C.L.;
P.C.M., T.Q.W. and Z.H.L. performed the ARPES measurements and analyzed the data;
H.C.L., Z.H.L. and W.X.L. wrote the manuscript based on discussion with all the authors.

\item[Supplementary Information] accompanies this paper.

\item[Author Information] The authors declare no competing financial interests. The data that support the findings of this study are available from the corresponding
author H.C.L. (hlei@ruc.edu.cn) upon reasonable request.

%\item[Correspondence] Correspondence and requests for materials should be addressed to Hongming Weng (email: hmweng@iphy.ac.cn), Shancai Wang (email: scw@ruc.edu.cn) or to Hechang Lei (email: hlei@ruc.edu.cn).

\end{addendum}

\newpage

\begin{figure}
%\centerline{\includegraphics[width=0.8\columnwidth]{fig1}} \vspace*{-0.3cm}
  \centerline{\epsfig{figure=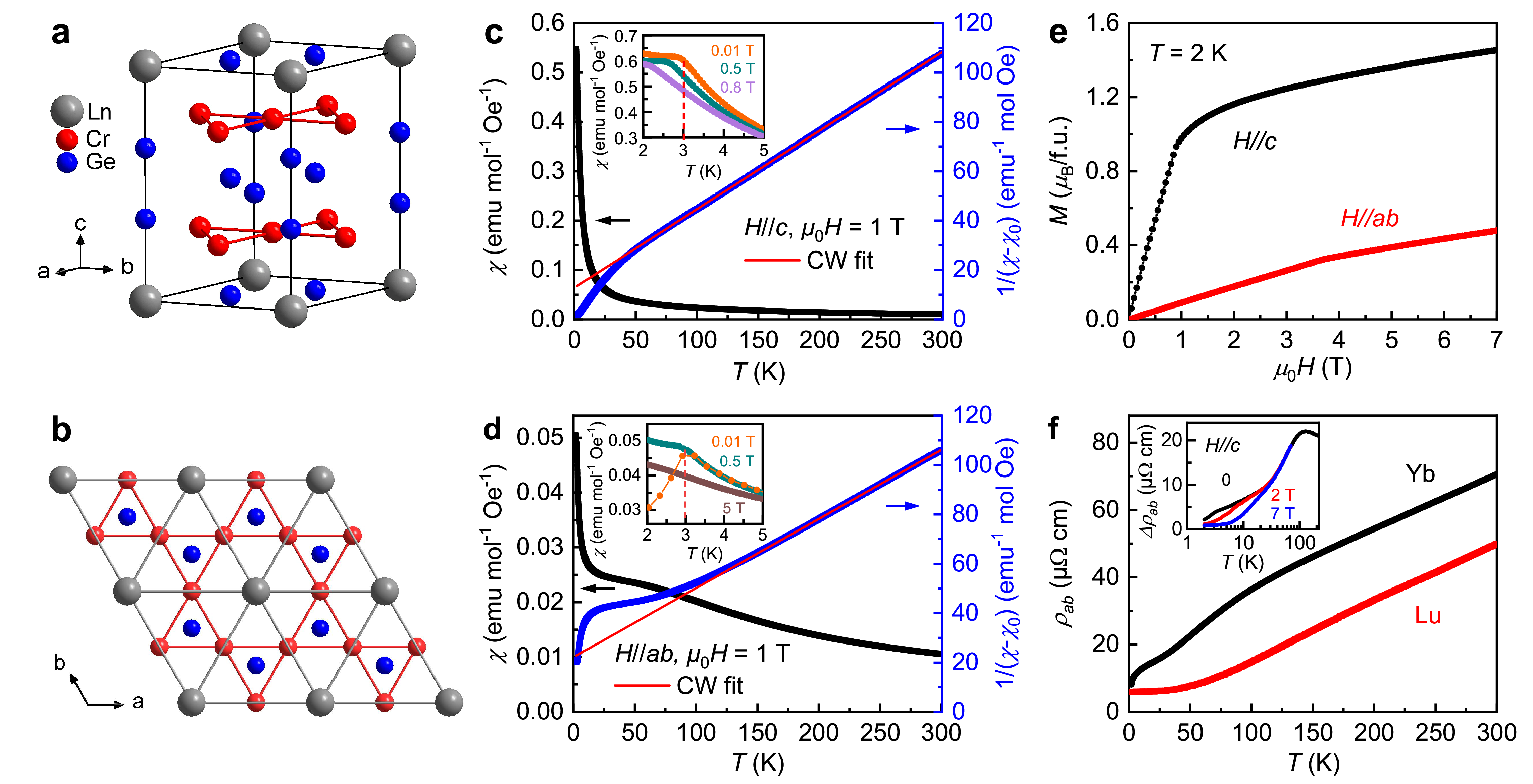,width=1.1\columnwidth}}
  \caption{\textbf{Crystal structure, magnetization and resistivity of LnCr$_{6}$Ge$_{6}$ single crystals.} 
    \textbf{a}, Crystal structure of LnCr$_{6}$Ge$_{6}$. The big gray, medium red and small blue balls represent Ln, Cr and Ge atoms, respectively. 
    \textbf{b}, Top view of LnCr$_{6}$Ge$_{6}$ crystal structure to highlight the Ln triangle lattice and Cr kagome lattice. 
    \textbf{c} and \textbf{d}, Temperature dependence of $\chi(T)$ (left axis) and $(\chi(T)-\chi_{0})^{-1}$(right axis) of YbCr$_{6}$Ge$_{6}$ single crystal at $\mu_{0} H$ = 1 T for $H\Vert c$ and $H\Vert ab$, respectively. The red solid lines represent the fits using the CW law at high-temperature region. Insets:  $\chi(T)$ curves at various fields around $T_{N}$.
    \textbf{e}, Field dependence of $M(\mu_{0}H)$ at $T$ = 2 K for both $H\Vert c$ (black symbols) and $H\Vert ab$ (red symbols). 
    \textbf{f}, zero-field $\rho_{ab}(T)$ as a function of temperature for YbCr$_{6}$Ge$_{6}$ and LuCr$_{6}$Ge$_{6}$ single crystals. Inset: $\Delta\rho_{ab}$ vs $T$ for YbCr$_{6}$Ge$_{6}$ from 1.8 K to 120 K at $\mu_{0} H$ = 0, 2 and 7 T.
 }
\end{figure}

\newpage

\begin{figure}
%\centerline{\includegraphics[width=0.8\columnwidth]{fig3}} \vspace*{-0.3cm}
  \centerline{\epsfig{figure=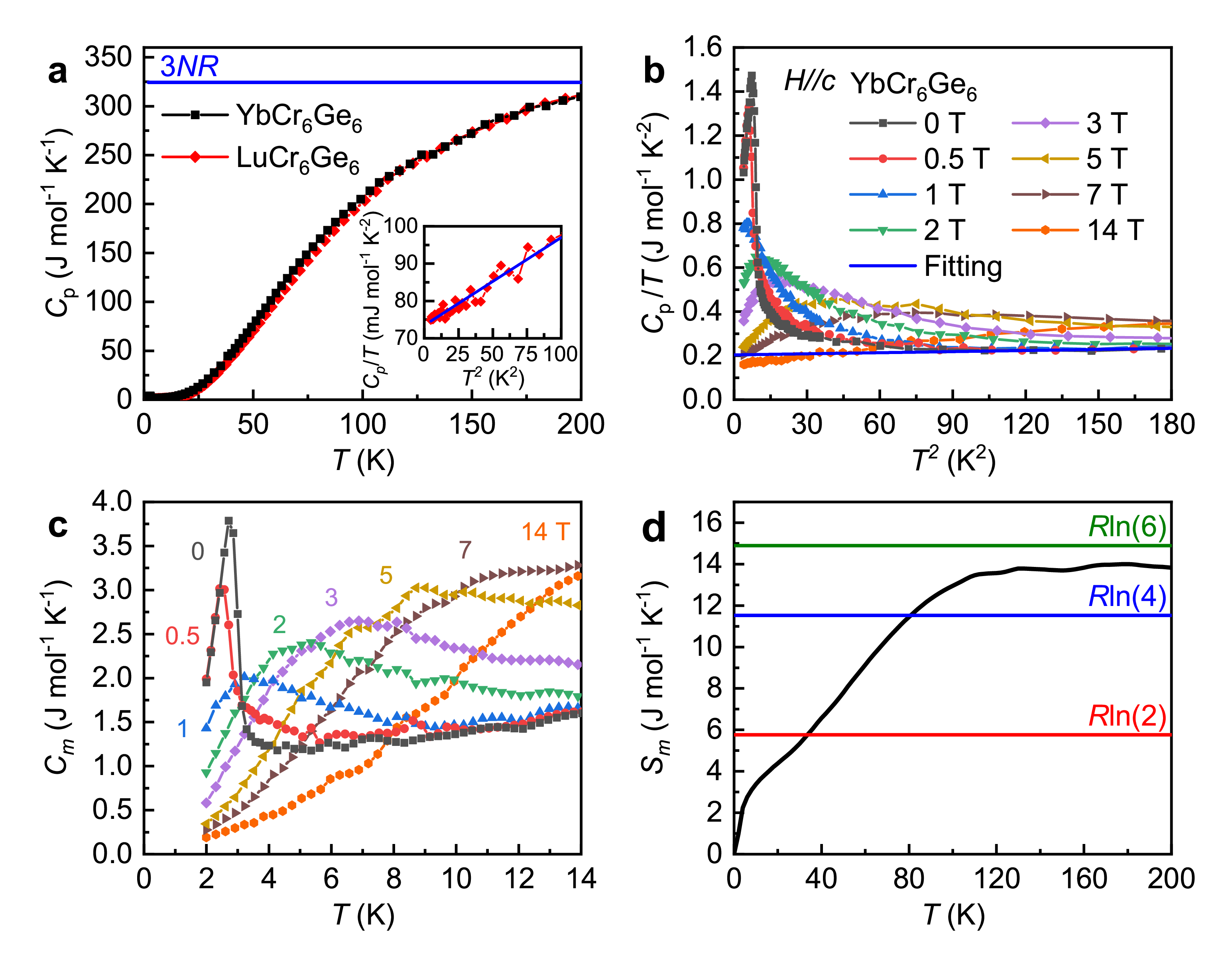,width=0.85\columnwidth}}
  \caption{\textbf{Specific heat and magnetic entropy.}
  	\textbf{a}, Temperature dependence of zero-field specific heat $C_p(T)$ between 2 K and  200 K for YbCr$_{6}$Ge$_{6}$ (black) and LuCr$_{6}$Ge$_{6}$ (red). Inset shows the linear fit of $C_{p}/T$ vs. $T^2$ for LuCr$_{6}$Ge$_{6}$.
  	\textbf{b}, Low-temperature $C_{p}/T$ of YbCr$_{6}$Ge$_{6}$ as a function of $T^2$ in different magnetic fields for $H\Vert c$.
  	\textbf{c}, Low-temperature magnetic specific heat $C_{m}(T)$ of YbCr$_{6}$Ge$_{6}$ in various magnetic fields, after subtracting the specific heat of LuCr$_{6}$Ge$_{6}$. 
  	\textbf{d}, Magnetic entropy $S_{m}(T)$ of YbCr$_{6}$Ge$_{6}$ in zero field.
  }
\end{figure}

\newpage

\begin{figure}
%\centerline{\includegraphics[width=0.8\columnwidth]{fig4}} \vspace*{-0.3cm}
  \centerline{\epsfig{figure=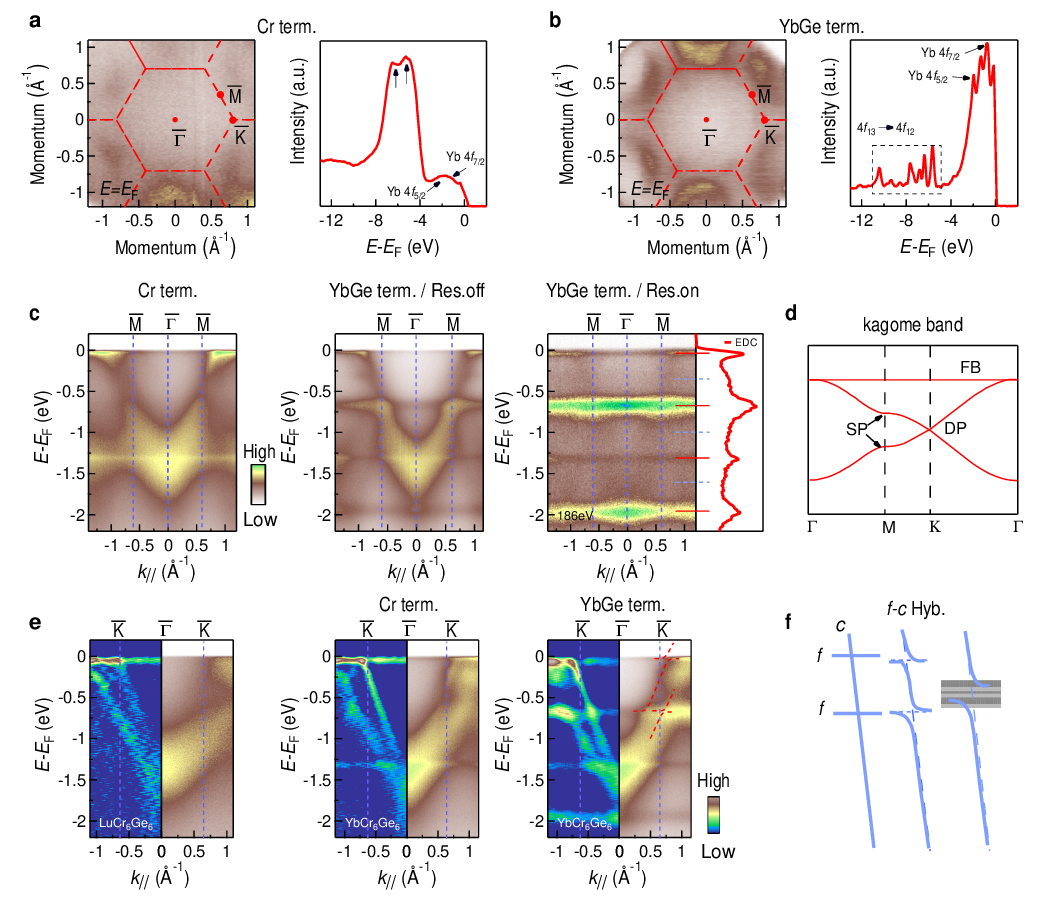,width=1\columnwidth}}
  \caption{\textbf{Termination-dependent Fermi surfaces and low-energy electronic structure.} 
  	\textbf{a} and \textbf{b}, Fermi surface maps and corresponding core-level spectra measured on the Cr- and YbGe-terminated surfaces of YbCr$_6$Ge$_6$, respectively. The red frames denote the 2D Brillouin zone projected onto the (001) plane. All data were acquired using a photon energy of 92 eV (corresponding to the $k_z \sim$ 0 plane), unless otherwise specified. 
  	\textbf{c}, ARPES intensity plots along the $\bar{\Gamma}$--$\bar{M}$ direction for the two surface terminations. The data shown in the right panel were taken using 186~eV photons, corresponding to the Yb $4d$--$4f$ resonant photoemission condition. The integrated EDC reveals nearly equally energy-spaced $4f$-derived peaks. 
  	\textbf{d}, Representative kagome band structure obtained from a tight-binding calculation, highlighting the flat bands, saddle points, and Dirac points. 
  	\textbf{e}, ARPES intensity plots and corresponding second-derivative maps along the $\bar{\Gamma}$--$\bar{K}$ direction for LuCr$_6$Ge$_6$ and for the two surface terminations of YbCr$_6$Ge$_6$. 
  	\textbf{f}, Schematic illustration of the hybridization between the conduction band and the localized Yb-$4f$ bands.}
\end{figure}

\newpage

\begin{figure}
%\centerline{\includegraphics[width=0.8\columnwidth]{fig4}} \vspace*{-0.3cm}
  \centerline{\epsfig{figure=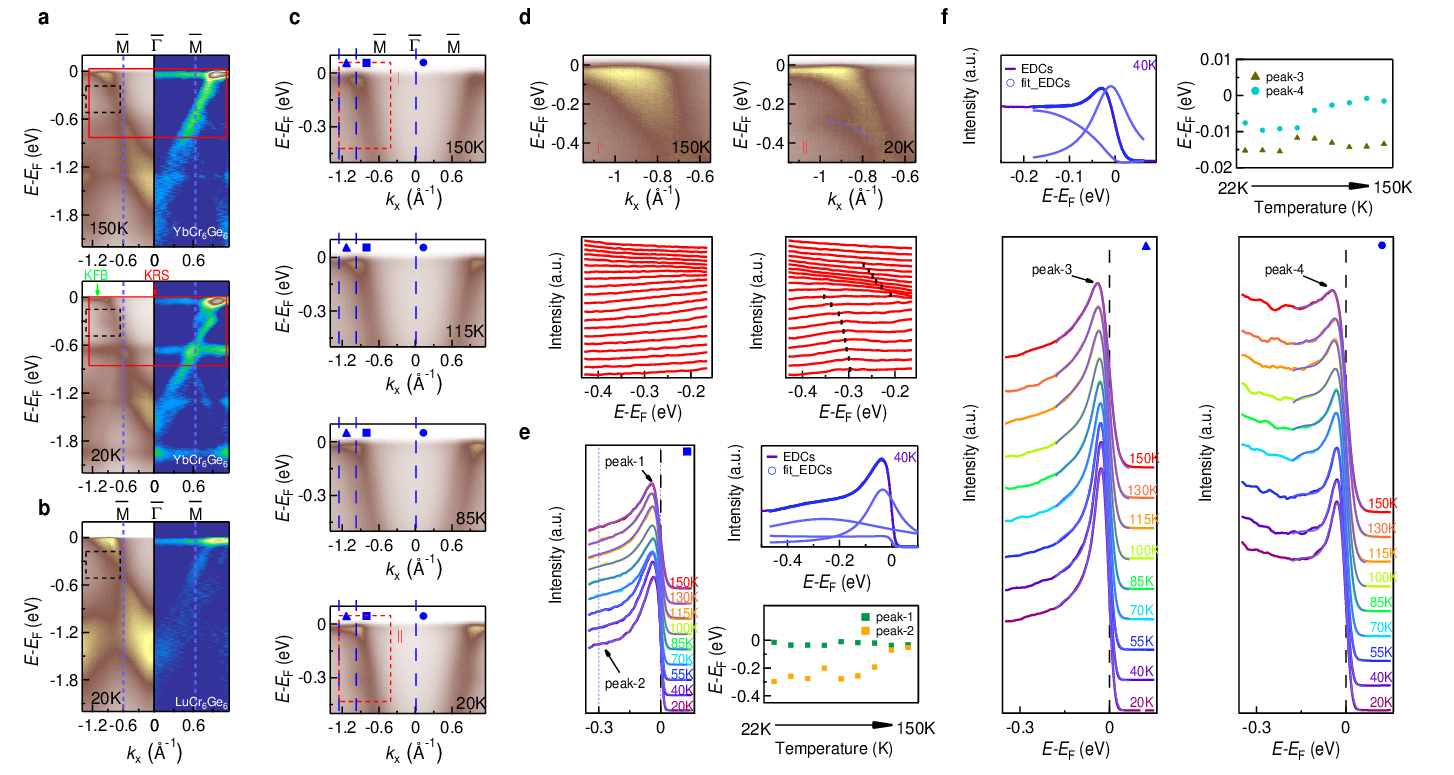,width=1.1\columnwidth}}
  \caption{\textbf{Temperature-dependent evolution of the flat bands.} 
	\textbf{a}, ARPES intensity plots and corresponding second-derivative maps along the $\bar{\Gamma}$--$\bar{M}$ direction on the $k_z \sim$ 0 plane for the YbGe-terminated surface of YbCr$_6$Ge$_6$, acquired at 150 K and 20 K, respectively. The green markers labeled KFB denote the kagome FBs, while the red markers labeled KRS indicate the Kondo resonance state.
	\textbf{b}, The same measurements as in \textbf{a}, but performed on LuCr$_6$Ge$_6$ at 20 K. 
	\textbf{c}, ARPES intensity plots of the enlarged regions in \textbf{a}, measured at 150 K, 115 K, 85 K, and 20 K, respectively. 
	\textbf{d}, Enlarged views of the red dashed boxes in \textbf{c} at 150 K and 20 K, highlighting clear temperature-induced changes. The corresponding EDCs are shown below. 
	\textbf{e}, Temperature evolution of EDCs at the momentum positions marked by the blue square in \textbf{c}. The EDCs are fitted using a polynomial background and Lorentzian functions. The extracted peak positions are summarized in the right panels, where the size of the squares represents the maximum fitting uncertainty. 
	\textbf{f}, Same analysis as in \textbf{e}, but for the momentum positions indicated by the blue triangle and blue circle in \textbf{c}.}
\end{figure}

\end{document}